\pdfoutput=1 

\documentclass[aps,prl,showpacs,twocolumn,floatfix,longbibliography]{revtex4-1}

\usepackage[utf8]{inputenc}
\usepackage{amsmath,amssymb,amsfonts}
\usepackage{graphicx,epsf}
\usepackage{xspace}
\usepackage{textcomp}
\usepackage{color}
\usepackage[normalem]{ulem}
\usepackage[caption=false]{subfig}
\usepackage{dsfont} 

\usepackage[colorlinks,bookmarks=false,citecolor=blue,linkcolor=blue,urlcolor=blue,hyperfootnotes=true]{hyperref}

\begin{document}

\title{Reconciling edge states with compressible stripes in a ballistic mesoscopic conductor}

\author{Pacome Armagnat}
\affiliation{Univ. Grenoble Alpes, CEA, IRIG-PHELIQS GT, F-38000 Grenoble, France}
\author{Xavier Waintal}
\affiliation{Univ. Grenoble Alpes, CEA, IRIG-PHELIQS GT, F-38000 Grenoble, France}

\begin{abstract}
The well-known Landauer-Buttiker (LB) picture used to explain the quantum Hall effect uses the concept of (chiral) edge states that carry the current. In their seminal 1992 article, Chklovskii, Shklovskii and Glazman (CSG) showed that the LB picture does not account for some very basic properties of the gas, such as its density profile, as it lacks a proper treatment of the electrostatic energy. They showed that, instead, one should consider alternated stripes of compressible and incompressible phases. In this letter, we revisit this issue using a full solution of the quantum-electrostatic problem of a narrow ballistic conductor, beyond the CSG approach. We recover the LB channels at low field and the CSG compressible/incompressible stripes at high field. Our calculations reveal the existence of a third "hybrid" phase at intermediate field. This hybrid phase has well defined LB type edge states, yet possesses a Landau level pinned at the Fermi energy as in the CSG picture. 
We calculate the magneto-conductance which reveals the interplay between the LB and CSG regimes. 
Our results have important implications for the propagation of edge magneto-plasmons.
\end{abstract}
\date{}
\maketitle

Electrostatic energy is very often the largest energy scale in a physical situation. Yet, the electrostatic landscape is
equally often taken for granted as an external potential, which may result in a wrong physical picture. 
A well known example is the quantum Hall effect\cite{Klitzing1980} (QHE) that has been largely discussed using the concept of edge states in a
non-interacting Landauer-Buttiker (LB) picture\cite{Halperin1982, Buettiker1988}. Despite being very successful for the understanding of e.g. the 
quantification of the Hall resistance and the vanishing longitudinal resistance, the LB picture also fails spectacularly to describe basic
physics such as the density profile of the electron gas. In a series of articles\cite{Wulf1988, Beenakker1990, Chang1990} that culminated with the work of Chklovskii, Shklovskii and Glazman (CSG)\cite{Chklovskii1992, Chklovskii1992a, Chklovskii1993}, it was recognized that the LB picture should be revisited.
It was shown that the interplay between quantum mechanics and electrostatics leads to the emergence of compressible and incompressible stripes, a concept related, yet somehow different, to the original edge states.
An important effort has been devoted to the experimental observation of these stripes
\cite{Schmerek1996, Haren1995, Wei1998, Aoki2005, Haren1995a, Panos2014, Patlatiuk2018, Kendirlik2017, Pascher2014, Ilani2004, Zhang2013}. 
CSG work, as well as a large fraction of the subsequent litterature\cite{Lier1994, Gueven2003, Oh1997, Siddiki2003, Siddiki2007, Bilgec2010, Gerhardts2013, Kavruk2011,Suzuki1998, Suzuki1996}  was based on the Thomas-Fermi approximation which is suitable at high magnetic field but inadequate at low field where the LB approach is expected to work well. 
More recent works improved on Thomas-Fermi by incorporating a Gaussian broadening of the Landau levels \cite{Siddiki2004,Salman2013,Sahasrabudhe2018}.
Solving the full self-consistent electrostatic-quantum problem, beyond the above approximations, is a difficult task however, as the presence of the Landau levels (and the associated Dirac comb for the density of states) makes the set of equations highly non-linear. In this letter, we use a newly developped numerical technique capable of handling this problem\cite{Armagnat2019} and explore how the LB channels present at low field evolve into CSG compressible stripes at high magnetic field. Using the solution of the full self-consistent problem,
we find that in a large region of the parameter space the system is in an "hybrid" phase that borrows features from both the LB and CSG pictures.

\begin{figure}[h!]
\centering
\includegraphics[width=\linewidth]{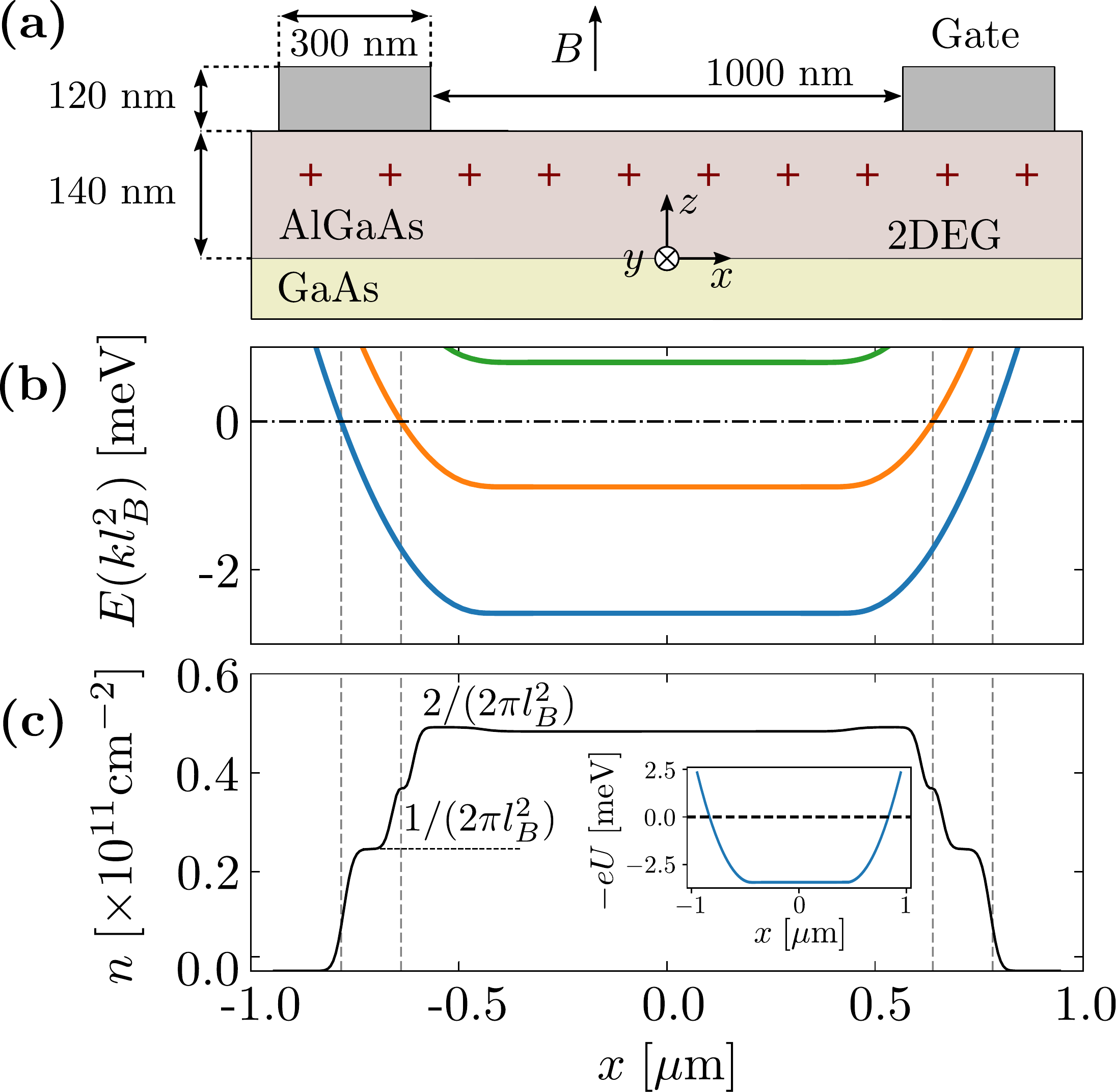}
\caption{Non-interacting picture of QHE. a) side view of the system (infinite along the $y$ direction). b) Dispersion relation 
$E_n(x=kl_B^2)$ for the three lowest Landau levels. c) density profile $n(x)$ (thin line) with Fermi level $E_F=0$ and $B= 1T$. The results of b) and c) have been calculated using a direct numerical solution of Eq.(\ref{eq:2deg_quantum}) with the external potential $U(x)$ shown in the inset of c).}
\label{fig:schema_2deg}
\end{figure}

\begin{figure*}[ht]
\centering
\includegraphics[width=\linewidth]{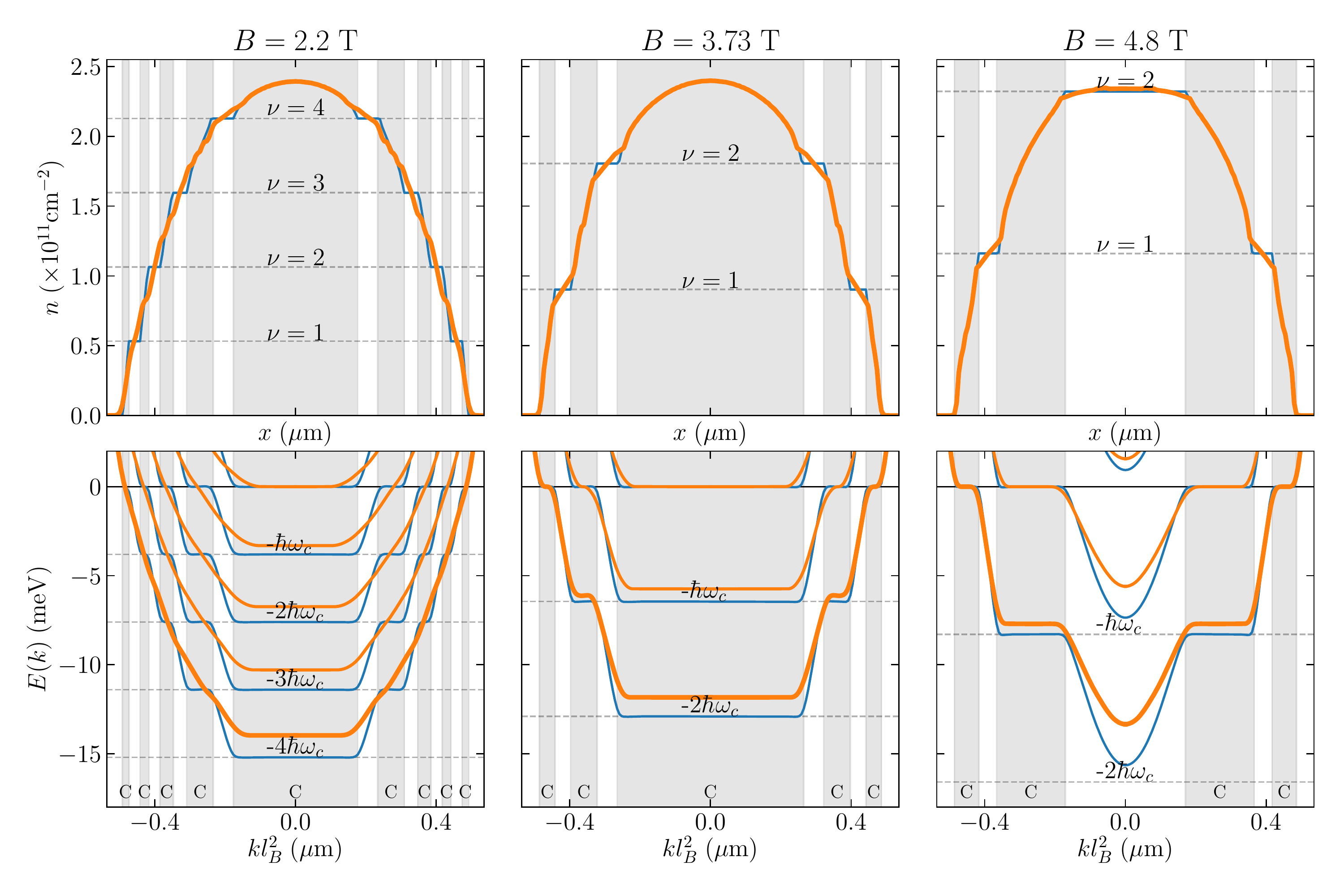}
\caption{Electronic density $n(x)$ (top) and band structure $E_n(k)$ (bottom) for three different values of the magnetic field at confinement $V_g=-0.75$V. Blue line: self-consistent Thomas-Fermi calculation; orange lines: full self consistent solution of the quantum-Poisson problem Eqs.(\ref{eq:2deg_quantum}), (\ref{eq:full_Poisson}), (\ref{eq:discont}) and (\ref{eq:full_density}). The gray "C" regions indicate the compressible stripes while the white regions are incompressible.}

\label{fig:6panels}
\end{figure*}
{\it The LB picture of the QHE regime (and its failure)}.  We consider the infinite wire geometry of Fig.\ref{fig:schema_2deg}a: a two dimensional electron gas at the interface between GaAs and GaAlAs is placed under a perpendicular magnetic field $B$ and confined to quasi-1D through two gates situated a few tens of nm above the gas. 
In the effective mass approximation, the electronic wave function $\psi(x,y)$ is described by a simple Schr\"odinger equation,
\begin{equation}
\label{eq:2deg_quantum}
\frac{1}{2 m^\star} \left( i \hbar \vec{\nabla} - e\vec A \right)^2 \psi(x,y) - e U(x) \psi(x,y) = E \psi(x,y)
\end{equation}
where the vector potential takes the form $\vec A = B x \vec y$ (Landau gauge) and $U(x)$ is the electrostatic potential ($m^*$: effective masss, $e$: electron charge). In the LB picture, the function $U(x)$ is an external input of the problem.
The general solution of Eq.(\ref{eq:2deg_quantum}) takes the form of plane waves $\psi_n(x,y)=\psi_n(x) e^{iky}$ along the $y$-direction
with momentum $k$.
In the absence of $U(x)$, these states are simply the Landau levels\cite{Landau1981}: equally spaced highly degenerate dispersiveless levels $E_n(k) = \hbar \omega_c (n+ 1/2)$ with states $\psi_n(x)$ that are exponentially localized along the $x$-direction around $x_k = k l_B^2$ ($l_B = \sqrt{\hbar / (e B)}$: magnetic length, $\omega_c = e B / m^\star$: cyclotron frequency, $n$ integer). 
Applying an external confining potential $U(x)$ around the edges of the sample (such as the one shown in the inset of Fig.\ref{fig:schema_2deg}c) provides the usual Landauer-Buttiker (LB) picture for the edge states of the QHE. For a slowly varying $U(x)$ (Thomas-Fermi approximation), the spectrum becomes dispersive and follows the potential
\begin{equation}
\label{eq:TF}
E_n(k) \approx \hbar \omega_c (n+ 1/2)-eU(x_k)
\end{equation} 
$E_n(k)$ crosses the Fermi energy ($E_F=0$) and provides conducting LB channels. The localized channels are the edge states of the system. The corresponding band structure is shown in Fig.\ref{fig:schema_2deg}b. These edge states are localized around $x_k$ and chiral (the velocity $v_k = (1/\hbar) dE/dk$ is positive for the edge states on the right of the sample and negative on the other edge). The associated
velocity is directly linked to the confining potential $v_k \approx (l_B^2/\hbar) dU(x)/dx$.

The main problem with the LB picture can be seen in Fig.\ref{fig:schema_2deg}c) where the associated electronic density profile $n(x)$ is shown. In the bulk of the system there are exactly two filled Landau levels, hence the electronic density is given by their corresponding degeneracy $n(x)=2/(2 \pi l_B^2) = 2eB/h$. 
As one moves towards the edge of the sample, one reaches the point where the second Landau level is not filled any more
and the density drops to $n(x)=1/(2 \pi l_B^2)$ and eventually to $n(x)=0$. In this picture, the density is essentially set by the magnetic field $B$.
However, the typical energy associated with the field $\hbar\omega_c$ is of the order of 10meV which is several order of magnitude smaller than
the electrostatic energy that would be needed to deform whatever electronic density $n(X,B=0)$ was there at $B=0$ into the one of 
Fig.\ref{fig:schema_2deg}c). One concludes that the potential $U(x)$ should not be considered as an external input but rather as
the solution of the Poisson equation,
\begin{equation}
\label{eq:full_Poisson}
 \Delta U(\vec{r}) = -\frac{e}{\epsilon}  n^{\rm{d}}(\vec{r})
\end{equation}
\begin{equation}
\label{eq:discont}
\frac{dU}{dz}(z=0^+)-\frac{dU}{dz}(z=0^-)= -\frac{e}{\epsilon} n(x)
\end{equation}
where $n^{\rm{d}}(\vec{r})$ is the (3D) dopant density and $n(x)$ the (2D) density of the electronic gas. In our wire geometry, 
the electronic density is given by
\begin{equation}
\label{eq:full_density}
n(x) =\sum_n \int \frac{dk}{2\pi} \left| \psi_{nk}(x)\right|^2 f(E)
\end{equation}
which closes our system of equation ($f(E)=1/(e^{E/k_BT}+1)$ is the Fermi function). In the QHE regime, Eqs.(\ref{eq:2deg_quantum}), (\ref{eq:full_Poisson}), (\ref{eq:discont}) and (\ref{eq:full_density}) form a highly non-linear set of equations.

{\it The CSG picture of compressible and incompressible stripes.} The strength of CSG argument is that the entire physical picture can be constructed from simple considerations, essentially from the dispersion relation Eq.(\ref{eq:TF}) that relates the energy of the Landau level 
$E_n(k)$ to its spatial position $x_k$. Let us suppose that we know the density profile $n(x,B=0)$ in the absence of magnetic field. 
We further assume that this profile (which results from the interplay between the electrostatic and kinetic energy) is only weakly affected by the presence of the magnetic field (as argued above the cyclotron energy $\hbar\omega_c$ is small compared to the electrostatic energy needed to strongly modify the density profile). For a generic value of the field, it follows that the filling fraction 
$\nu \equiv n_0 2 \pi l_B^2$ of the Landau levels in the bulk of the sample ($n_0=n(x=0,B=0)$  has generically a non-integer value e.g. $\nu= 2.4$. From that statement, it follows that a 
Landau level (in this example the third one) {\it lies exactly at} the Fermi energy since it is only partially occupied. This is a very different situation from the non-interacting picture discussed above. As one gets away from the center of the sample towards the edge, the filling factor  $\nu (x) \equiv n(x,B=0) 2 \pi l_B^2$ decreases until the third Landau level is totally depleted and one starts to deplete the second Landau level.
Depleting the second Landau level implies that it sits at the Fermi energy, hence a sharp rise of the electrostatic energy (of amplitude $\hbar\omega_c$). In the small region where this sharp rise occurs, the density is constant (no available level at the Fermi energy). This region is an incompressible stripe. Continuing towards the edge, we hence obtain a set of compressible stripes separated by incompressible stripes. The blue lines of Fig.\ref{fig:6panels} shows the resulting dispersion relation (lower panels) and density profile (upper panels) for
three different value of the magnetic field. The blue lines of Fig.\ref{fig:6panels} ressemble very strongly the cartoon shown in Fig.1 of the CSG paper. However, they correspond to a full self-consistent calculation in the Thomas-Fermi approximation.

{\it Hybrid phase at intermediate fields.} 
We now turn
to the full self-consistent solution of the problem Eqs.(\ref{eq:2deg_quantum}), (\ref{eq:full_Poisson}), (\ref{eq:discont}) and (\ref{eq:full_density}) without using the Thomas-Fermi approximation. Note that in what follows, the Landau levels are supposed to be fully spin polarized, we do not discuss the magnetic instabilities.
The results are shown with orange lines
in Fig.\ref{fig:6panels}. At high field (middle and right panels), the full solution bears strong similarities with the Thomas-Fermi result 
and one gets alternated compressible and incompressible stripes with the middle of the sample being compressible (middle panels) or
incompressible (right panel, corresponds to a quantized plateau of conductance) \cite{Chklovskii1993}. One qualitative differences
is the absence of well defined plateaus of the density in the incompressible region. This is due to the fact that the Landau levels spread over
a width $l_B$ which is not infinitely small compared to the width of the incompressible stripes (e.g. $l_B \approx 13$ nm at $B=3.73$ T).
At very low magnetic fields (not shown), one fully recover the LB picture with well defined propagating channels that cross the Fermi level.
We find that the transition between the LB picture and the CSG one at high field happens in two stages: first, the formation of Landau levels
that get pinned at the Fermi level; secondly the evolution of the edge states into compressible stripes. In the corresponding intermediate field range, the system is in an intermediate "hybrid" phase with well defined edge states 
(similar to those shown in Fig.\ref{fig:schema_2deg}) yet with
a central compressible stripe that remains pinned at the Fermi level. This situation is illustrated in the left panel of 
Fig.\ref{fig:6panels} ($B=2.2$ T, corresponding to $\nu\approx 4.5$). 
        
{\it Magneto-conductance of ballistic wires.} To gain further insight, we now turn to a discussion of the current that flows upon applying a small bias voltage across the wire. Note that the question of where does the current flow is ill defined in the Thomas-Fermi approximation, at least at very small temperature. Indeed the current must flow in the compressible regions since transport requires available states at the Fermi level. Yet in the compressible regions, the velocity $v_k = (1/\hbar) dE/dk$ vanishes, hence one would expect the current to do the same.
This small paradox (which points to the current being concentrated to the edge of the compressible stripes) is resolved by using the full self-consistent solution.
\begin{figure}[ht]
\centering
\includegraphics[width=\linewidth]{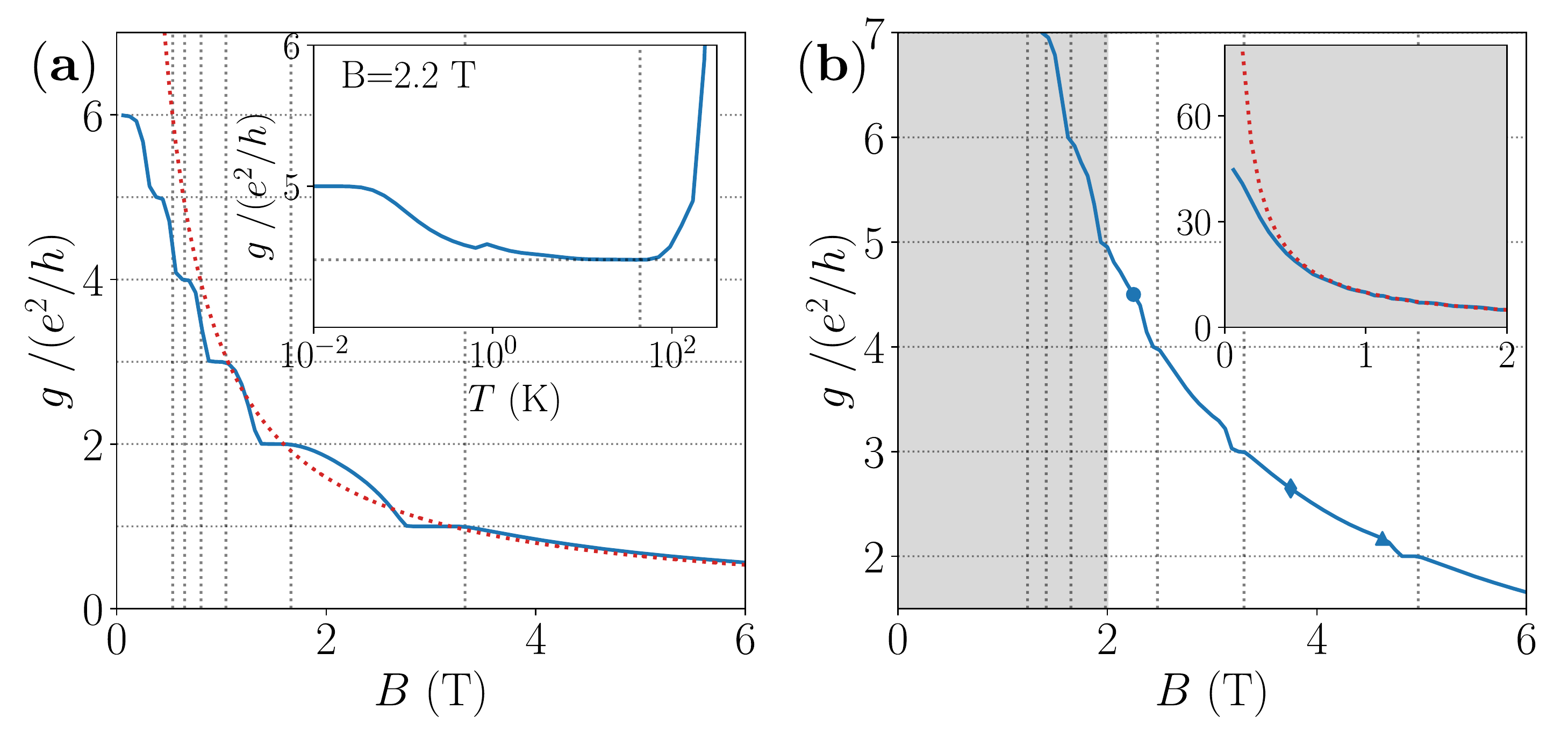}
	\caption{{\bf (a)} Conductance $g(B) $as a function of the magnetic field for a confinement $V_g = -1.6$V at $T=1$K (thin blue line). The dotted line indicates the $g = n_0 e/ B$ law, horizontal lines indicate quantized values of the conductance while the vertical lines the expected position of the Hall plateaus from the bulk density $n_0$: $B_\nu = n_0 h/(e \nu)$. Inset: $g(T)$ as a function of the temperature (log scale) for a confinement $V_g=-0.75$ V. The vertical line indicates $k_BT =\hbar \omega_c$. 
{\bf (b)} Same as (a) for $V_g = -0.75$ V. The round, diamond and triangle symbols mark respectively the conductance at $B=2.2, 3.73 \text{ and }4.8 $T of Fig.\ref{fig:6panels}. Inset: zoom of the main panel at small field.}
\label{fig:conductance_TandB}
\end{figure}
In a perfectly ballistic system where all conducting channels are perfectly transmitted, the conductance $g$ is given by a simple form of the Landauer formula,
\begin{equation}
\label{eq:conductance}
g = e^2 \sum_n \int \frac{\rm{d}k}{2 \pi} \; 
\theta(v_{nk}) v_{nk} \left(-\frac{\partial f}{\partial E}[E_n(k)]\right) 
\end{equation}
where the Heaviside function $\theta (x)$ selects the channels with positive velocities. At zero temperature, this formula provides
the well-known quantization of conductance: using $v_k = (1/\hbar) dE/dk$ and the fact that $df/dE\rightarrow -\delta(E)$ one finds that
$g$ simply counts the number of bands that cross the Fermi level (in unit of $e^2/h$). The CSG situation where there is a degenerate band exactly at the Fermi level is a new situation as it can lead to {\it non-quantized} conductance even in a ballistic conductor. Indeed, assuming that $E_n(k)$ is a (very slowly) increasing function of $k$ from $k=0$ to $k=\infty$, one gets $g = (e^2/h) \sum_n f[E(k=0)] - f[E(k=\infty)]$
which translates into $g = (e^2/h) \nu$. Since the central stripe is in general not fully filled, this results in a 
non quantized conductance that scales as $\sim 1/B$ except for the plateaus occuring when the central region is incompressible. This situation corresponds to a regime where the temperature is very small compared to the cyclotron frequency, yet large compared to the small variations of the electric potential in the central Landau level. It is in sharp contrast with the zero temperature limit where quantization is always expected.
 
This is illustrated in the inset of Fig.\ref{fig:conductance_TandB}a where we plot the conductance versus temperature in the hybrid phase: We identify three regimes: a regime of large temperature (with respect to $\hbar\omega_c$) and accordingly large conductance; a regime of low temperature where one recovers the LB quantization; and an intermediate regime without quantization  $g=(e^2/h) \nu(B)$. The crossover between the later two is strongly system dependent: it depends on the small curvature of the electric field in the bulk of the sample, hence on the strength of the gate voltage and magnetic field. It is interesting to note that even for moderate confinement (here $V_g=-0.75$V and the gas still occupies most of the space in between the two electrodes, this crossover takes place at $T\sim 100 mK$, i.e. much higher than typical dilution fridge temperatures. 
The transition from the LB to the CSG regime can also be seen as a function of magnetic field, see 
Fig.\ref{fig:conductance_TandB}a (strong confinement) and Fig.\ref{fig:conductance_TandB}b (weak confinement). At low field, one observes the LB plateaus of conductance which
are replaced at higher field by a general $1/B$ law entertwined with the Hall plateaus. The crossover between the two regimes can be identified from the deviation of the conductance with respect to the
$g=(e^2/h) \nu(B)= n_0 e/ B$ law (dotted line). Note that the $g=n_0 e/ B$ law can be extracted directly from the experiment since the bulk density $n_0$ can be extracted from the $B$-position 
$B_n = n_0 h/(e n)$ ($n=1,2\dots$) of a Hall plateau at high field. 

{\it Discussion.} The full self-consistent approach of the quantum-electrostatic problem allows one to treat on the same footing the low field LB regime and the large field CSG regime. Likewize, it allows to understand how one crossovers from well defined conducting channels at zero temperature to the CSG quantum Hall regime. Our most surprising result is that the quantum Hall effect appears in two stages which leads to the hybrid phase described above. A rough estimate of the parametric regime where this hybrid phase is expected goes as follows. The typical width $w$ of a compressible stripe is set by the density profile $n(x,B=0)$. If $x_0$ defines the position of an integer filling $\nu=n(x_0,B=0) 2 \pi l_B^2$, then  $n(x_0+w,B=0) 2 \pi l_B^2= \nu-1$. Noting $d$ the typical distance with which $n(x_0,B=0)$ falls from its bulk value to zero, we get $w\sim (d/\nu)$.
$d$ depends strongly on the electrostatics of the problem; it is of the order of the distance between the
gate and the gas, $d\sim 140 nm$ in our example. The crossover from the compressible stripe behaviour to the LB like edge state is expected when quantum fluctuations are large enough, i.e. $w\sim l_B$. This translates into $\nu_0\sim (n_0 d^2)^{1/3}$. Hence, for filling
fraction larger than $\nu_0\sim (n_0 d^2)^{1/3}$ one expects LB like edge states while for higher magnetic field, one recovers the CSG stripes. In our (rather typical) example, the hybrid scenario corresponds to most Hall plateaus $\nu>(n_0 d^2)^{1/3}\approx 3$. 

As the LB edge states and CSG stripes are both associated with one quantum of conductance, it remains to
discuss the difference between the two situations in actual observables. We expect an important difference in the propagation of voltage pulses \cite{Kamata2010} (edge magneto-plasmons).
Indeed, the plasmon velocity is proportional to the Fermi velocity (up to a renormalization factor due to
Coulomb interaction\cite{Matveev1993, Roussely2018}) of the corresponding mode. For a LB edge state,
this velocity $v_k = (1/\hbar) dE/dk$ is a well defined quantity that depends on the confining potential
i.e. $v_k\sim \hbar/(dm^*)$.
We find typical values $v_k\sim 10^5$ m.s$^{-1}$ consistent with the values quoted in the litterature.
The situation is drastically different for a CSG compressible stripe where the velocity vanishes in the middle of the stripe and sharply rises on its boundaries. If follows that the average velocity in the stripe drops down upon entering the CSG regime. Perhaps more importantly, the velocity now strongly depends on $k$ which results in an important spreading of the excitation between the slow part in the middle of the stripe and the faster part toward its edges. Hence, we expect that at high field, voltage pulses will get highly distorded in sharp contrast with the behavior in the hybrid phase at lower field. We anticipate that the hybrid phase is more favorable for the propagation of pulses than the CSG phase. 

{\it Aknowledgement.} This work was supported by the ANR Full Quantum, the ANR QTERA and the US Office of Naval Research. We thank F. Portier, P. Roche and C. Glattli for usefull discussions.

\bibliographystyle{naturemag}
\bibliography{qhe_revisited_biblio}

\end{document}